\documentclass[sigplan,screen]{acmart}
\AtBeginDocument{%
  }

\usepackage{booktabs}
\usepackage{longtable}
\usepackage{array}
\usepackage{ragged2e}
\usepackage{multirow}
\usepackage{tikz}
 \usetikzlibrary{positioning, arrows.meta, calc, backgrounds}
\newcolumntype{L}[1]{>{\RaggedRight\arraybackslash}p{#1}}
\setcopyright{cc}
\setcctype{by}
\acmDOI{10.1145/3840586.3843744}
\acmYear{2026}
\copyrightyear{2026}
\acmISBN{979-8-4007-2934-8/2026/10}
\acmConference[Onward! '26]{Proceedings of the 2026 ACM SIGPLAN International Symposium on New Ideas, New Paradigms, and Reflections on Programming and Software}{October 4--9, 2026}{Oakland, CA, USA}
\acmBooktitle{Proceedings of the 2026 ACM SIGPLAN International Symposium on New Ideas, New Paradigms, and Reflections on Programming and Software (Onward! '26), October 4--9, 2026, Oakland, CA, USA}
\acmSubmissionID{onward26essays-p17-p}
\received{2026-05-15}
\received[accepted]{2026-06-28}
\begin{document}
\title{Agents as Software: A Programming Languages Agenda for Agent Reliability}
\author{Shraddha Barke}
\orcid{0009-0000-4573-9296}
\affiliation{%
  \institution{Microsoft Research}
  \city{Redmond}
  \country{USA}
}
\email{sbarke@microsoft.com}

\author{Adithya Murali}
\orcid{0000-0002-6311-1467}
\affiliation{%
  \institution{University of Wisconsin-Madison}
  \city{Madison}
  \country{USA}
}
\email{adithyamurali@cs.wisc.edu}
\begin{abstract}
AI agents increasingly resemble software systems: they call tools, remember facts, follow policies, delegate work, and take actions with real consequences.
Yet the ``program'' of an agent is scattered across prompts, tools, memories, workflows, and execution traces, making its behavior difficult to inspect through ordinary testing and debugging alone.
This essay argues that a programming-systems perspective offers a natural lens for making agents reliable.
We recast agents as programmable artifacts whose behavior can be specified over traces and state, checked before deployment, monitored during execution, and improved from observed failures.
The goal is not to make probabilistic agents behave like deterministic programs, but to give them enough structure that their
behavior can be reasoned about, controlled, and repaired.
\end{abstract}
\ccsdesc[300]{Software and its engineering~Formal methods}
\ccsdesc[300]{Computing methodologies~Intelligent agents}

\keywords{agent reliability, specifications, logics for agents, static and dynamic analysis, agent verification}
\maketitle

\section{Introduction}
AI agents are emerging as the next major abstraction in software engineering: systems capable of reasoning over natural language and acting through software interfaces.
Agents can be viewed as probabilistic microservices: modular entities that perform specialized tasks, but with underspecified constraints and inherent randomness. 
Unlike deterministic software systems, agents operate in dynamic, uncertain environments, interpreting ambiguous user intent while taking actions that may have real-world side effects.
From a programming languages perspective, the central shift is that the agent's ``program'' is no longer contained in a single body of source code. Its behavior is distributed across prompts, tools, memories, policies, plans, traces, and orchestration logic.
Agents are therefore programmable systems, but their programming model is implicit, probabilistic, and only partially specified. 
This makes conventional testing and debugging inadequate.

Ensuring the correctness and reliability of such systems therefore demands a new discipline for specifying, analyzing, monitoring, and improving agent behavior. 
This paper argues for a programming languages view of AI agents. 
Rather than treating agents as opaque wrappers around language models, we view them as software artifacts whose distributed behavior can be studied using the tools of programming languages, formal methods, runtime verification, and neuro-symbolic reasoning.
This view starts with defining specifications for agents. 
Specifications can describe what counts as valid behavior over event logs, state abstractions, permissions, information flow, tool preconditions, belief updates, and resource usage. 
They can express familiar software properties, such as type correctness, resource discipline, and authorization, while also capturing agent-specific concerns such as grounded belief, stale evidence, tool-use justification, and recoverability after failure.
%
%

Of course, specifications must also be connected to analyses that operate before, during, and after execution. 
Static analysis techniques treat prompts, plans, tool schemas, memory policies, and configuration artifacts as analyzable objects before deployment, checking whether the agent is structurally capable of reliable behavior.
This includes checking for ambiguous obligations, invalid data and tool flows, and high-impact actions that become reachable before their required evidence has been established.

Dynamic analysis complements these checks by tracking what actually happens during execution.
By maintaining an evolving state of goals, tool outputs, evidence, permissions, obligations, and constraints, a runtime monitor can check whether actions are grounded, beliefs remain valid, irreversible operations are authorized, and global trace properties continue to hold.
%
%
%
Finally, execution traces can become material for learning better specifications, repairing workflows, optimizing tool use, and evolving agents as their environments change.
In this sense, agent reliability is not a single verification problem, but an ongoing systems problem: how to design agents that can be checked before execution, monitored during execution, and improved after execution. 
The goal is not to eliminate the probabilistic nature of agents, but to make their behavior more explicit, inspectable, and governable.
Determinism is not the goal: sampling is an implementation mechanism, in the way a thread scheduler or a garbage collector is, and software has long presented predictable interfaces over such mechanisms by constraining what they are allowed to do rather than by making them deterministic.

Much of what follows is not agent-specific: these ideas are well established in programming languages and systems research, and should be adapted rather than reinvented.
Three assumptions, however, do not carry over cleanly.
First, the artifact being analyzed is not necessarily fixed before execution: an agent may revise its memory, plans, instructions, or generated code in ways that alter its subsequent behavior.
Second, atomic predicates need not have deterministic interpretations: whether food is \emph{rotten}, for example, may be decided by a fallible model rather than a fixed procedure.
Third, an execution need not admit a canonical abstraction: the same run may be projected into several event logs, and monitors over those logs may disagree on whether the specification was violated.

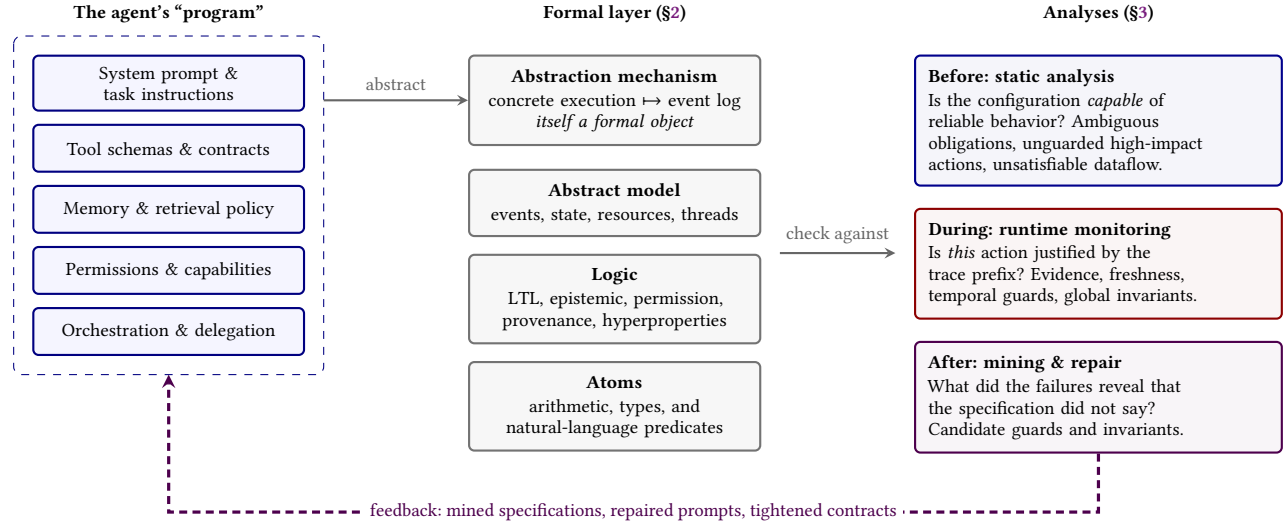
\begin{figure*}[t]
\centering
\begin{tikzpicture}[
    >=stealth,
    font=\small,
    artifact/.style={
        rectangle, rounded corners=2pt, draw=blue!50!black, thick,
        fill=blue!4, text width=3.3cm, align=center, inner sep=4pt,
        font=\scriptsize, minimum height=0.62cm
    },
    middlebox/.style={
        rectangle, rounded corners=2pt, draw=black!55, thick,
        fill=black!3, text width=3.5cm, align=center, inner sep=5pt,
        font=\scriptsize
    },
    phasebox/.style={
        rectangle, rounded corners=2pt, draw=#1, thick,
        fill=#1!4, text width=4.5cm, align=left, inner sep=5pt,
        font=\scriptsize
    },
    grouplabel/.style={font=\scriptsize\bfseries, align=center},
]

\node[grouplabel, anchor=north] (l1) at (0, 0) {The agent's ``program''};

\node[artifact, anchor=north] (a1) at ([yshift=-0.30cm]l1.south) {System prompt \& task instructions};
\node[artifact, anchor=north] (a2) at ([yshift=-0.16cm]a1.south) {Tool schemas \& contracts};
\node[artifact, anchor=north] (a3) at ([yshift=-0.16cm]a2.south) {Memory \& retrieval policy};
\node[artifact, anchor=north] (a4) at ([yshift=-0.16cm]a3.south) {Permissions \& capabilities};
\node[artifact, anchor=north] (a5) at ([yshift=-0.16cm]a4.south) {Orchestration \& delegation};

\draw[draw=blue!50!black, dashed, rounded corners=3pt]
  ([xshift=-7pt, yshift=7pt]a1.north west) rectangle ([xshift=7pt, yshift=-7pt]a5.south east);

\node[grouplabel, anchor=north] (l2) at (5.9, 0) {Formal layer (\S\ref{sec:specifications})};

\node[middlebox, anchor=north] (abs) at ([yshift=-0.30cm]l2.south) {%
    \textbf{Abstraction mechanism}\\[1pt]
    concrete execution $\mapsto$ event log\\
    \emph{itself a formal object}%
};

\node[middlebox, anchor=north] (model) at ([yshift=-0.30cm]abs.south) {%
    \textbf{Abstract model}\\[1pt]
    events, state, resources, threads%
};

\node[middlebox, anchor=north] (logic) at ([yshift=-0.22cm]model.south) {%
    \textbf{Logic}\\[1pt]
    LTL, epistemic, permission,\\ provenance, hyperproperties%
};

\node[middlebox, anchor=north] (atoms) at ([yshift=-0.22cm]logic.south) {%
    \textbf{Atoms}\\[1pt]
    arithmetic, types, and\\ natural-language predicates%
};

\node[grouplabel, anchor=north] (l3) at (12.3, 0) {Analyses (\S\ref{sec:analysis})};

\node[phasebox=blue!55!black, anchor=north] (static) at ([yshift=-0.30cm]l3.south) {%
    {\bfseries Before: static analysis}\\[1pt]
    Is the configuration \emph{capable} of\\ reliable behavior? Ambiguous\\ obligations, unguarded high-impact\\ actions, unsatisfiable dataflow.%
};

\node[phasebox=red!55!black, anchor=north] (dynamic) at ([yshift=-0.28cm]static.south) {%
    {\bfseries During: runtime monitoring}\\[1pt]
    Is \emph{this} action justified by the\\ trace prefix? Evidence, freshness,\\ temporal guards, global invariants.%
};

\node[phasebox=violet!60!black, anchor=north] (post) at ([yshift=-0.28cm]dynamic.south) {%
    {\bfseries After: mining \& repair}\\[1pt]
    What did the failures reveal that\\ the specification did not say?\\ Candidate guards and invariants.%
};


\draw[->, thick, black!60]
  ([xshift=7pt]a1.east |- abs) -- node[above, font=\scriptsize] {abstract} (abs.west);

\coordinate (fmid) at ($(abs.north)!0.5!(atoms.south)$);
\draw[->, thick, black!60]
  ([xshift=7pt]abs.east |- fmid) -- node[above, font=\scriptsize] {check against}
  ([xshift=-7pt]static.west |- fmid);

\path let \p1 = (a5.south), \p2 = (atoms.south), \p3 = (post.south) in
  coordinate (base) at (0, {min(\y1, min(\y2, \y3)) - 22});

\draw[->, very thick, violet!60!black, densely dashed]
  (post.south) -- (post.south |- base)
  -- node[font=\scriptsize, text=violet!60!black, fill=white, inner sep=2pt]
       {feedback: mined specifications, repaired prompts, tightened contracts}
     (a5.south |- base)
  -- ([yshift=-7pt]a5.south);

\end{tikzpicture}
\caption{The programming-systems view of an agent. The agent's ``program'' (left)
is distributed across artifacts in different languages, some of which are written
during execution. An abstraction mechanism lifts concrete executions into event
traces, over which specifications are stated in some logic and vocabulary of atoms
(middle). Analyses (right) consume these before, during, and after execution, and
what they observe is fed back into the artifacts.}
\label{fig:overview}
\end{figure*}

The rest of this paper develops this perspective. Section~\ref{sec:specifications} discusses specifications for AI agents, including abstract models of agentic computation, logics for stating agent properties, domain-specific predicates expressed through natural language or neuro-symbolic mechanisms, and the role of language design in making such specifications easier to enforce.
Section~\ref{sec:analysis} then turns to static and dynamic analysis, showing how prompts, tools, memory, retrieval, and execution traces can be checked for reliability properties.
Figure~\ref{fig:overview} summarizes how the pieces fit together.

\section{Specifications for AI Agents}
\label{sec:specifications}
A science of reliability must start with a discussion of risks, and the specifications that we can enforce to mitigate them: What do we want to prevent agentic systems from doing? How do we formulate good behavior?

In this section, we will essentially stay close to the classical programming languages understanding of bad behavior in software: memory mishandling, permission violations, leaked secrets, inputs/outputs of the wrong type, functional incorrectness, etc, which can be defined as computational problems. 
There is a significant body of literature on "harms" that AI systems can perpetuate that is centered more on humanistic and social concerns~\cite{harm-incidents-survey,harm-reduction-taxonomy,deepmind-sociotechnical-safetyeval}.
While some of these harms may be expressible as formal specifications (e.g., privacy or fairness), this is not true in general (e.g., preventing generation of graphic images, whatever that may mean in a specific cultural context). 
Additionally, we will dedicate our discussion to the problem of stating formal specifications and leave the discussion of how they may be enforced to later sections.

We conceive the science of specifications for agents as consisting of three pieces: (1) an abstract model that formalizes agentic computation, (2) a compositional logic that specifies the family of properties we wish to state --- such that the formulas in the logic can be interpreted over the chosen abstract model, and (3) a set of atomic functions and predicates that encapsulate useful aspects of the domain at hand and over which the specification formulas will be written. 


\paragraph{Formalizing Specifications} The first step towards formalizing specifications is to define a formal computational \emph{model} of agents. One can then use various logics, languages, and calculi to express specifications over the abstract model. 
We observe that agentic frameworks are closer to systems software, since their computation is often best understood as an open-ended process consisting of individual atomic steps.
Agentic frameworks also feature complex concurrency, delegation protocols, and reactivity in their design, which reinforces the aptness of this comparison. 
In this spirit, we propose that agentic executions can be abstracted into a log of ``events''.
The vocabulary of the events tracked in the agents' computation provides a specific abstraction, since two agents that are indistinguishable at the level of a particular event vocabulary would satisfy the same properties. 
Therefore, the choice of what events must be logged can be left up to the designer of the verification infrastructure (human or otherwise). 

In this essay, we do not take a position on specific modeling choices, such as choosing between action-based and state-based formalisms~\cite{csp-hoare1978,csp-book,pi-calculus-book,tla-lamport94}. 
At the time of writing this essay, popular discussions on agent reliability tend to exemplify specifications that are more naturally stated over state-like abstractions, e.g., permission tokens for accessing files, resource invariants (like number of sub-agents allowed), or well-typed-ness of values. 
State-based formalisms like TLA+~\cite{tla-lamport94,LamportSpecifyingSystems} also offer robust tool support for working with such specifications, and can therefore be a more attractive choice if applicable. 
However, it may be difficult to commit to purely state-based formalisms for dynamic agents, since agents can decompose tasks in different ways and may not admit a single cohesive abstract state that captures an agent's execution.
For example, consider a code agent that spawns sub-agents to refactor related functionality in a code repository into one or more functions.
The sub-agents may traverse and modify the repository in different orders, making it difficult to identify a single abstract state invariant that should hold across all intermediate executions.
In contrast, modeling the changes as an event log where agents commit changes they make to the log, respecting function call order or other hierarchies in the code can be more useful. 
Action-based formalisms can therefore offer better support for stating and composing higher-level coordination specifications for complex agent workflows.



Unsurprisingly, some emerging works on agentic specifications advocate for adopting an event-based abstraction of execution behavior~\cite{agentspec,agentguard}. 
Many of these works predominantly focus on violations involving improper actions in certain states/environments (e.g., improper tool use)~\cite{agentspec,agentguard,odersky-scala-capabilites-agents,kamath2025enforcing,guardians-of-agents}, and agents are therefore modeled as executing a sequence of well-defined (and often predefined) actions. 
They then utilize the boundaries between LLMs and tools or initial/terminal points to define events, e.g., after an action has been completed by an agent, or when a tool communicates a result.

There is a richer space of modeling choices for such abstractions. 
For instance, one can model tools and the resources at an agent's disposal and define stringent interfaces between different agents and tools. 
One can explicitly model memory and have a vocabulary over memory stores and accesses performed by agents, or model the environment and use vocabulary over the environment's responses. 
Another natural extension is to have a formal model that distinguishes sub-agents with their own virtual `thread'. 
One interesting realization of this model is to require certain events to be committed to a concrete log, potentially in the middle of an agent's execution.

A subtlety that can arise with event-based abstractions--- as opposed to formalisms that only model agent-tool-human boundaries--- is that we are no longer guaranteed that a given concrete trace can only be mapped on to a single abstract trace. 
For example, consider an abstraction that records an $\mathit{acquire}(\mathit{agent\_id},r)$ and a $\mathit{release}(\mathit{agent\_id},r)$ for a resource $r$, with the specification being that a resource cannot be acquired twice without being released in the middle. 
Now consider a trace where an agent delegates a sub-agent to operate on a resource it acquired. 
Depending on how the trace is abstracted, one may record the release of the resource by the parent agent either before the sub-agent begins or after the sub-agent's execution has finished. Not only are these distinct abstractions, but one of them satisfies the specification while the other does not! Therefore, the abstraction mechanism for lifting real executions to abstract traces must be carefully defined (and treated as a formal object in its own right) when working with richer abstract models.

Finally, we observe that an event log-based view is only an abstraction and inherits the weaknesses of incompleteness that plagues any abstraction mechanism. For example, given a vocabulary consisting of sending and receiving emails, an agent may respect the specification ``read email from ABC $\rightarrow$ send response to ABC if ABC is not a no-reply address'' by behaving as follows: ``read email $\rightarrow$ leak password $\rightarrow$ send email''. The password leak will not be detected by a mechanism enforcing the given specification. This is obviously because we haven't even modeled that leaking passwords is bad. One can mitigate such issues to some extent by using other specification frameworks that attempt to model the environment explicitly and prevent certain actions outright, or using a default-restrictive as opposed to a default-permissive model--- meaning that any actions beyond benign ones must be modeled explicitly to be allowed. An interesting spin on this approach is the idea of ``capability permissions'', which are permission resources that allow an agent to only exhibit certain classes of capabilities.
For instance, in the above example, one can say that no actions beyond reading and responding to emails are allowed. A system enforcing such a specification can then flag the password leak as being outside the scope of capabilities allowed for the agent. More generally, a security-oriented treatment of agentic reliability may need to consider more involved abstraction and specification mechanisms. We do not investigate this further in this essay.

\begin{table*}[t]
\centering
\caption{An example set of agent specifications, varied by specification purpose and type of logic that may be used to express the property. The Usage column notes whether these properties are typically found in a survey of related work on agent reliability (R), were not found in related work by the authors verbatim but are essentially natural extensions of those currently used (B as in Boundary), or are speculated by the authors (S).}
\label{tab:agent-specs}
\footnotesize
\begin{tabular}{@{}L{6cm} L{3.4cm} L{4.5cm} c@{}}
\toprule
\textbf{Spec example} & \textbf{Property kind} & \textbf{Logic} & \textbf{Usage} \\
\midrule

Tool inputs/outputs must satisfy the given pre/post- condition
& Interface correctness
& FOL
& R \\

Every \texttt{Acquire(r)} must have a matching \texttt{Release(r)} later before another \texttt{Acquire(r)}
& Resource well-formedness
& Regexes over trace vocabulary
& R \\
 
Cannot issue \texttt{delete\_resource(r)} without prior \texttt{Owns(agent, r)} in session
& Action admissibility
& LTL
& R \\
 
Order retrieval must be preceded by an authentication of the corresponding customer
& Action admissibility
& LTL
& R \\
 
\texttt{confirmation\_token} argument equals token from prior \texttt{request\_confirmation}; not reused
& Input integrity (freshness / use-once)
& Session types
& R \\
 
Capabilities held by subagent $\subseteq$ capabilities held by parent at spawn
& Permission discipline
& Permission Logics
& R \\
 
No use of a capability after its \texttt{Revoke} event
& Permission discipline
& LTL
& R \\
 
No \texttt{Send\_external(m)} where \texttt{m} derives from a \texttt{confidential}-labeled source
& Information-flow (label-based)
& Information flow types/logics
& R \\
 
Output independent of content from demarcated potential junk source
& Counterfactual property/Hyperproperty
& HyperLTL or similar logics
& B \\
 
No tool invocation with arguments derived from \texttt{untrusted} data without calling \texttt{validate}
& Value provenance
& FOL/Gradual Types/Provenance Logics
& B \\
 
Every output supported by a corresponding appropriate \texttt{Source} 
& Output faithfulness (anti-hallucination)
& Provenance logics
& B \\
 
Tool call expenditures within remaining budget
& Quantitative
& Trace logics with counting/Arithmetic constraints
& R \\
 
Probability of reaching unsafe state $\leq \varepsilon$
& Quantitative + Safety
& PCTL / probabilistic LTL
& R \\
 
Secret input and observable output share $\leq n$ bits
& Quantitative
& FOL with Arithmetic
& B \\
 
Every irreversible action (e.g., tool call) is derivable from the action set of an \texttt{Authorized\_Plan(p)}
& Plan-coherence/ provenance
& LTL + Provenance/Permission logics
& B \\

Before any irreversible action, the agent's belief in the justification for the action is derivable from prior plans in the trace
& Beliefs and Justification
& Epistemic logics
& B/S \\
 
No subagent acting for user $u$ reads data labeled for user $u' \neq u$
& Data security/Access management
& FOL/Permission logics
& B \\
 
Subagents spawned in disjoint scopes do not share resources
& Access management
& Resource logics e.g., separation logic
& B/S \\
 
Events measured over user-level time (e.g., must produce result within wall-clock $n$ seconds)
& Realtime properties
& Timed logics
& B \\

An agent does not continue to hold a belief after it has seen evidence to contradict the belief
& Belief consistency
& Dynamic Epistemic logics
& S \\

Cooperating subagents must have mutual knowledge of a shared plan
& Multi-agent coordination
& Epistemic logics over multiple parties
& S \\
\bottomrule
\end{tabular}
\end{table*}
\paragraph{Logics for Agentic Specifications} Once we fix an abstract model, we can express specifications as logical formulas that are evaluated over the model. There are abundantly many choices here.  In our view, the key concerns are not raw expressiveness or elegance from a purely specification perspective: anyone who simply wants to convey a property can do so in plain English, and if someone knows how a property can be checked they can simply encode that check directly as a program over the model. 
Instead, the considerations are more about the algorithmic toolkit associated with the logics. 
Different logics vary in their support for automated deduction, the complexity of deciding whether a property holds, and the extent to which specifications can be composed.
We discuss some techniques for static and dynamic analysis for various classes of properties in later sections. 

Table~\ref{tab:agent-specs} illustrates an assortment of properties that one may wish to employ for describing reliable behavior in agents. Many of these are of course familiar to readers acquainted with verification and testing of systems software. Among these, the class of resource and permission-based specifications appears to be the most interesting and immediate challenge to tackle. Since agents require complex coordination and utilize dynamic delegation to divide tasks, reasoning about whether an individual agent instance has permission to access/modify a resource or call a tool, and in particular being able to maintain such reasoning across an unbounded trace by monitoring provenance of permission tokens appears to be a challenging and useful problem for coordinating agents at scale.

An interesting class of properties that is less explored in contemporary literature is knowledge and belief logics. 
One of the most attractive aspects of learning-based systems is their ability to act under incomplete information based on learned knowledge of the domain, and therefore one of the risks is that an agent acts on knowledge it does not actually have, or on beliefs that may be contradicted by subsequent evidence. 
This class of specifications is slightly different from the kinds of permission specifications discussed above. Rather, they pertain to whether the agent knew or justifiably believed something at a particular point. We showcase several interesting variations on such specifications in Table~\ref{tab:agent-specs}. 

The simplest example is a grounded-belief specification, which asks that any belief acted upon must be derivable from the agent's observable history (whether that means the context, the memory, or perhaps tool outputs) and blocking actions which cannot be justified in this manner. 
As we discuss later under evidence-grounded tool use in Section~\ref{sec:runtime-analysis}, this kind of specification can be monitored dynamically by checking whether a proposed action is justified by facts derivable from the current trace prefix.
A revision-consistent specification requires the agent not to act on a belief once later evidence in the same trace has contradicted it. 
This blocks the failure mode where an agent latches onto an early conclusion and ignores subsequent evidence.
A coordination specification, in the multi-agent setting, asks that two agents not jointly commit to a plan unless each knows the plan and each knows the other knows it, blocking handoff failures where a planner and an executor disconnect. 
Each of these scenarios corresponds to concrete logics studied in the literature. Utilizing the language of knowledge and belief within a formal system to reason about agents' actions is a promising avenue for future work.
\paragraph{Atoms} The final piece in writing specifications is the set of atomic concepts, such as functions and predicates, that are determined by the domain of interest.
On one end, one can use standard mathematical and logical domains: integers or reals with arithmetic operators, sets and maps, booleans, and so on.
On the other end, we can simply choose to express atomic concepts in natural language, e.g. ``busy intersection'', which can be used to specify safe behavior of an autonomous vehicle,  or ``rotten food'', which can be used in the context of a cooking agent. 

Recent work in this area allows users to specify a general set of predicates in natural language and automates checking them either by prompting LLMs to evaluate these predicates on traces in a wholesale manner, or by asking LLMs to write Boolean programs (derived from the models' internal knowledge about the domain) which implement the predicate~\cite{agentspec,alchemist}. An interesting middle ground here has been explored by prior work on neuro-symbolic programming where neural `modules' capturing e.g., whether an image is that of a cat, are composed with programmatic or logical combinators~\cite{vipergpt,ijcai2022visualpuzzles,delta-fol,deepproblog,scallop}. Such frameworks offer the ability to enforce type systems on expressions in the language, learning distribution-specific semantics, and differentiable reasoning. 

However, what remains underdeveloped is a formal account of \emph{natural-language theories}: collections of predicates specified in natural language, evaluated by learned models, and connected by explicit relations such as implication, incompatibility, and revision.
This problem has clear predecessors.
Cyc encoded commonsense knowledge as axioms~\cite{lenat1995cyc}; ontology engineering developed shared vocabularies~\cite{gruber1995ontology}; description logics identified decidable fragments~\cite{dl-handbook}; the Semantic Web extended these ideas to the open web~\cite{semantic-web-revisited}; Wikidata curated knowledge at scale~\cite{wikidata}; and Wikifunctions now pursues their composable-function counterpart~\cite{falk2026wikilambda}.
These efforts provide much of the representational machinery we need.
New work should therefore reuse existing formalisms and vocabularies.
Language models change the economics: predicates are now cheap to evaluate but difficult to trust.
There are three aspects of such theories that remain to be developed.
First, one can investigate different kinds of natural language theories and how they may be formalized. For example, consider a predicate $\mathit{busy\_intersection}(arg)$. 
This predicate has a highly contextual interpretation: what it means for an intersection to be busy during work traffic is not the same as what it means at 2 a.m., and the meaning may also differ depending on whether the intersection is in a large city or a small town. 
One way to formalize such a predicate is to treat its meaning as context-dependent rather than fixed. For instance, \(\mathit{busy\_intersection}\) may be interpreted probabilistically, together with a complement predicate \(\mathit{not\_busy}\), so that their probabilities sum to one.
Alternatively, its semantics may be expressed in terms of confidence, possibility, or defeasible rules that can be revised when new evidence becomes available. 
We are not aware of a general framework that integrates these
reasoning mechanisms over fallible learned predicates.

Another aspect of these natural language theories is how various predicates may be related to each other, including any (hard or soft) entailments between them~\cite{semantic-regex-jocelyn}. For example, one would presumably want $\mathit{busy\_intersection}$ to entail $\mathit{yield\_before\_turn}$, or want $\mathit{rotten\_food}$ to be incompatible with $\mathit{safe\_to\_eat}$. 
This is also the point where mechanisms whose combinators are fixed \emph{a priori}---including logic programming---can find limitations, since there are innumerable ways to restate the same concept and natural language predicates can be related to each other in tricky ways.
We also know of no work that provides a well-considered formal footing for recording these entailments. Finally, developing standalone solvers for interesting and useful natural theories that can also be combined into larger reasoning mechanisms is an area that is completely unexplored. 

The ideal goal here is to develop abstract mechanisms for designing natural language theories with associated recipes for defining the collection of predicate signatures involved, the kinds of entailments or other relationships between them, and algorithms that can perform reasoning modulo these theories. We imagine this line of research to be similar to designing type systems, where related reasoning patterns are abstracted into higher level logical categories with associated reasoning algorithms.
A careful study of natural language theories is thus fertile ground for future work and would greatly benefit the science of agent reliability.
\noindent
We conclude this section with a brief note on designing languages for reliable agent engineering.

\paragraph{The Role of Language Design} Many specifications described above can be thought of as enforcing certain disciplines in agentic execution at an algorithmic level, in the sense that the specifications can already be stated at the level of the vocabulary of the task and/or agent orchestration.
Therefore, it is useful to explore prompt- and agent-programming languages that provide rich interfaces for integrating data types, permissions, and other reliability artifacts, such as test cases, into the specification of agent workflows.
Designing such languages, as with traditional programming languages, can reduce the burden of checking for certain properties during execution. This is essentially the ``correct-by-construction'' position on engineering reliability.

The technical questions in this area arise at two levels.
First, there are higher-level choices about the harnesses within which agents are deployed: whether agents are treated as relatively free actors, or whether they are only used to generate code that ultimately acts on their behalf after being subjected to formal verification~\cite{odersky-scala-capabilites-agents,guardians-of-agents}.
Second, there are design choices about the language abstractions used to write agents themselves: differing in the styles of traditional programming languages they choose to adopt or extend, their support for control flow, parallelization, or exposure to lower-level systems knobs for controlling LLMs (like context window, temperature, low-level prompting or decoding strategies) within the language, and their use of different constraint mechanisms for eliciting specific outputs from language models~\cite{lmql10.1145/3591300,pdlvaziri2024pdldeclarativepromptprogramming,dspykhattab2024,appldong2024,genaiscript2025,sglang,oracle-open-agent-spec}.
\section{Program Analysis for Agent Reliability}
\label{sec:analysis}
The previous section described how agent behavior can be specified over abstract models, logics, and domain-specific predicates.
We now ask how such specifications can be used to improve reliability.
We separate analysis along the execution boundary: static and dynamic analysis.
Together, these analyses support reliability before and during execution, while traces provide feedback for improving future agent behavior.

\subsection{Static Analysis}
Static analysis checks the artifacts that exist before an agent runs: prompts, tool schemas, policies, memory and retrieval configurations, and workflow structure.
Its goal is to identify configurations that permit unreliable behavior before they are deployed~\cite{cousot1977abstract,nielson1999principles,barke2026agentrx,10.1145/3786335.3813153}.

\paragraph{Agent Configurations as Static Artifacts.}
Static analysis for agents provides a way to reason about whether an agent is configured to behave reliably before it is executed.
Unlike a conventional single-program artifact, agentic behavior is distributed across prompts, tool schemas, policies, memory, and orchestration logic.
%
Static analysis treats these configuration artifacts as the objects to be checked.
The goal is not to predict the agent's exact execution path, but to check whether the system has been set up in a way that makes reliable execution more likely: are the instructions clear, are tool preconditions stated, are dangerous actions properly guarded, are data dependencies satisfiable, and can the available tools support the intended workflow?
A prompt, a tool schema, a retrieval policy, or a memory rule may each appear individually harmless, but their interaction can also cause failures.
For example, a prompt may instruct the agent to issue refunds, while the available tools do not expose whether a returned item has actually been received.
Similarly, a workflow may require user confirmation before an irreversible action, but the prompt may never state when such confirmation is required.
Static analysis makes these gaps visible before the agent is deployed.

\paragraph{Prompts as Interfaces.}

Prompts are the most visible configuration artifact in an agentic system, but they should not be treated as isolated text.
A prompt functions like an interface between the task, the model, the available tools, and the surrounding control policy.
Static analysis of prompts should therefore ask whether the prompt states the obligations needed for reliable execution: which instructions are trusted, when the agent should ask for clarification, when it should call a tool, when it should stop or escalate, whether irreversible actions require confirmation, and what output format downstream components expect~\cite{sharma2025promptpex,rzig2025empirically,zhu2023promptrobust}.

This makes prompt analysis less about stylistic prompt engineering and more about making prompts checkable.
Ambiguous directives such as ``resolve the issue'' or ``fetch relevant data'' are risky because they leave critical decisions underspecified: what evidence is sufficient, which tools may be used, what constraints must be preserved, and what actions require approval.
A static checker can flag missing placeholders, contradictory prompt sections, underspecified escalation rules, and outputs that do not conform to the JSON schema, DSL, or intermediate representation expected by downstream components.
In this sense, prompts become closer to executable interfaces: they expose obligations that can be checked against tools, policies, and workflows.
This mirrors a familiar pattern from program synthesis and compiler analysis, where partial programs, sketches, specifications, and intermediate representations are analyzed before execution to rule out ill-formed or unsafe completions~\cite{solar2005programming,gulwani2017program}.

\paragraph{Tool and Workflow Validation.}
Prompts describe what the agent should do, while tool schemas and tool-use workflows describe what actions are available and when those actions are valid.
Here, a tool-use workflow means the intended structure of an agent task: which tools may be called, what facts must be established before each call, and which actions require confirmation or authorization.
The goal of static analysis would be to check whether this structure is possible and safe before the agent runs~\cite{xavier2026agentproof}.
We illustrate the main checks using a refund or exchange workflow.

An agent should not be able to call \texttt{issue\_refund} merely because the user requested a refund; the workflow should establish that the order exists, the item is eligible, the return has been received, the refund amount is valid, and any required confirmation has been obtained.
Ensuring this involves three related checks: local action validity, dataflow validity, and workflow reachability.
The first check is local action validity.
At the simplest level, the tool call proposed by the model must have the right `shape': a valid tool name, required arguments, and schema-conformant values.
A more complex requirement would be similar to refinement types~\cite{vazou2014refinement}, which can go beyond shape, 
e.g.,  \texttt{amount > 0} and \texttt{amount <= order\_total(order\_id)}.
More advanced checks could involve looking at effects: 
reading a state or constructing an internal plan vs. mutating an account, contacting a user, deleting data, or moving money~\cite{odersky-scala-capabilites-agents}.

The second check is dataflow validity.
Tool arguments should be grounded in information that the agent actually has access to: user input, prior tool responses, retrieved documents, or authorized memory.
A checker can flag unresolved references, hallucinated identifiers, missing dependencies, and outputs that do not match the schema expected by later steps.
For example, an \texttt{exchange\_delivered\_order\_items} call should be traceable to a prior \texttt{get\_order\_details} call that established the relevant order, items, delivery status, and available replacement options.
Similarly, the analyzer can detect redundant or conflicting actions, such as workflows that make both \texttt{return\_item} and \texttt{exchange\_item} reachable for the same item without a disambiguating condition.

The third check is workflow reachability.
A locally valid action may still be unsafe if it becomes reachable too early.
For example, \texttt{issue\_refund} may have a valid \texttt{order\_id} and a valid \texttt{amount}, but it should not be reachable until the workflow has established that the order exists, the return has been received, and the refund amount is allowed.
An abstract-interpretation view is useful here: the analyzer can over-approximate possible agent states and flag paths where a high-impact action is reachable before its required evidence is available.
The warning is not that a particular execution has failed, but that the configuration permits a bad execution.
\begin{figure*}[t]
\centering
\begin{tikzpicture}[
    >=stealth,
    font=\small,
    passbox/.style={
        rectangle, rounded corners=2pt, draw=green!50!black, thick,
        fill=green!5, text width=6.5cm, align=left, inner sep=5pt
    },
    failbox/.style={
        rectangle, rounded corners=2pt, draw=red!60!black, thick,
        fill=red!5, text width=6.5cm, align=left, inner sep=5pt
    },
    replanbox/.style={
        rectangle, rounded corners=2pt, draw=violet!60!black, thick,
        fill=violet!5, text width=6.5cm, align=left, inner sep=5pt
    },
    statebox/.style={
        rectangle, rounded corners=2pt, draw=blue!50!black, thick,
        fill=blue!3, text width=6.8cm, align=left, inner sep=6pt
    },
    checkbox/.style={
        rectangle, rounded corners=2pt, draw=red!60!black, thick,
        fill=red!5, text width=6.8cm, align=left, inner sep=6pt
    },
    insightbox/.style={
        rectangle, rounded corners=2pt, draw=black!20, thick,
        fill=black!2, text width=14.8cm, align=left, inner sep=6pt
    },
    badge/.style={
        rectangle, rounded corners=2pt, fill=#1, text=white,
        font=\scriptsize\bfseries, inner sep=2.5pt
    },
]

\node[font=\small\bfseries, anchor=west] at (0, 0) {Agent Execution Trace};

\node[passbox, anchor=north west] (s1) at (0, -0.35) {%
    {\scriptsize\bfseries\color{green!40!black} Step 1}\\[1pt]
    \texttt{tool\_call(get\_order\_details,}\\
    \texttt{\quad \{id: 4821\})}%
};
\node[badge=green!50!black, anchor=north east]
    at ([xshift=-4pt, yshift=-4pt]s1.north east) {PASS};

\node[passbox, anchor=north west] (s2) at ([yshift=-0.35cm]s1.south west) {%
    {\scriptsize\bfseries\color{green!40!black} Step 2}\\[1pt]
    \texttt{tool\_call(check\_return\_policy,}\\
    \texttt{\quad \{order: 4821\})}%
};
\node[badge=green!50!black, anchor=north east]
    at ([xshift=-4pt, yshift=-4pt]s2.north east) {PASS};

\node[failbox, anchor=north west] (s3) at ([yshift=-0.35cm]s2.south west) {%
    {\scriptsize\bfseries\color{red!60!black} Step 3}\\[1pt]
    \texttt{tool\_call(issue\_refund,}\\
    \texttt{\quad \{order: 4821, amount: 127\})}\\[3pt]
    {\scriptsize\color{red!60!black}
    Required: \texttt{return\_status == "received"}\\
    Actual:\;\;\;\, \texttt{return\_status == "not\_shipped"}}%
};
\node[badge=red!60!black, anchor=north east]
    at ([xshift=-4pt, yshift=-4pt]s3.north east) {BLOCKED};

\node[replanbox, anchor=north west] (s4) at ([yshift=-0.35cm]s3.south west) {%
    {\scriptsize\bfseries\color{violet!60!black} Step 4 (re-plan)}\\[1pt]
    \texttt{tool\_call(request\_return\_confirmation,}\\
    \texttt{\quad \{order: 4821\})}%
};
\node[badge=violet!60!black, anchor=north east]
    at ([xshift=-4pt, yshift=-4pt]s4.north east) {SAFE};

\draw[->, thick, gray!50] (s1.south) -- (s2.north);
\draw[->, thick, gray!50] (s2.south) -- (s3.north);
\draw[->, thick, violet!50!black, densely dashed]
    (s3.south) --
    node[right, font=\scriptsize, text=violet!60!black, pos=0.4] {re-plan}
    (s4.north);

\node[font=\small\bfseries, anchor=west] at (8.2, 0) {Monitor State (after Step 2)};

\node[statebox, anchor=north west] (state) at (8.2, -0.35) {%
    \textbf{Established facts:}\\
    \quad \texttt{user\_authenticated = true}\\
    \quad \texttt{order\_exists(4821) = true}\\
    \quad \texttt{return\_status = "not\_shipped"}\\[4pt]
    \textbf{Pending obligations:}\\
    \quad \texttt{verify\_return\_receipt}\\
    \qquad \texttt{BEFORE issue\_refund}%
};

\node[checkbox, anchor=north west] (check) at ([yshift=-0.4cm]state.south west) {%
    {\bfseries\color{red!60!black}Precondition check}
    for \texttt{issue\_refund}:\\[3pt]
    \texttt{REQUIRED: return\_status == "received"}\\
    \texttt{ACTUAL:\;\;\;\, return\_status == "not\_shipped"}\\[3pt]
    {\bfseries\color{red!60!black}VERDICT: BLOCK.}
    Obligation not discharged.%
};

\draw[->, very thick, red!50!black, dashed]
    (s3.east) -- (check.west);

\node[insightbox, anchor=north west] (insight) at ([yshift=-0.5cm]s4.south west) {%
    \textbf{Key insight.}
    The monitor maintains a compact representation of the trace prefix: facts
    established so far, their provenance, constraints still in force, and
    unresolved obligations. For high-impact actions, the trace must show that
    required guards have been discharged in order; not merely that the relevant
    facts are present in the state.%
};

\end{tikzpicture}
\caption{Runtime monitor example. The agent attempts to issue a refund
(Step~3), but the monitor blocks the action because the required temporal
guard \texttt{verify\_return\_receipt} has not been discharged. The agent
then re-plans by first checking whether the return has been received.}
\Description{A runtime-monitoring example showing an agent trace, monitor state, a blocked refund action, and a safe re-planning step.}
\label{fig:runtime-monitor}
\end{figure*}
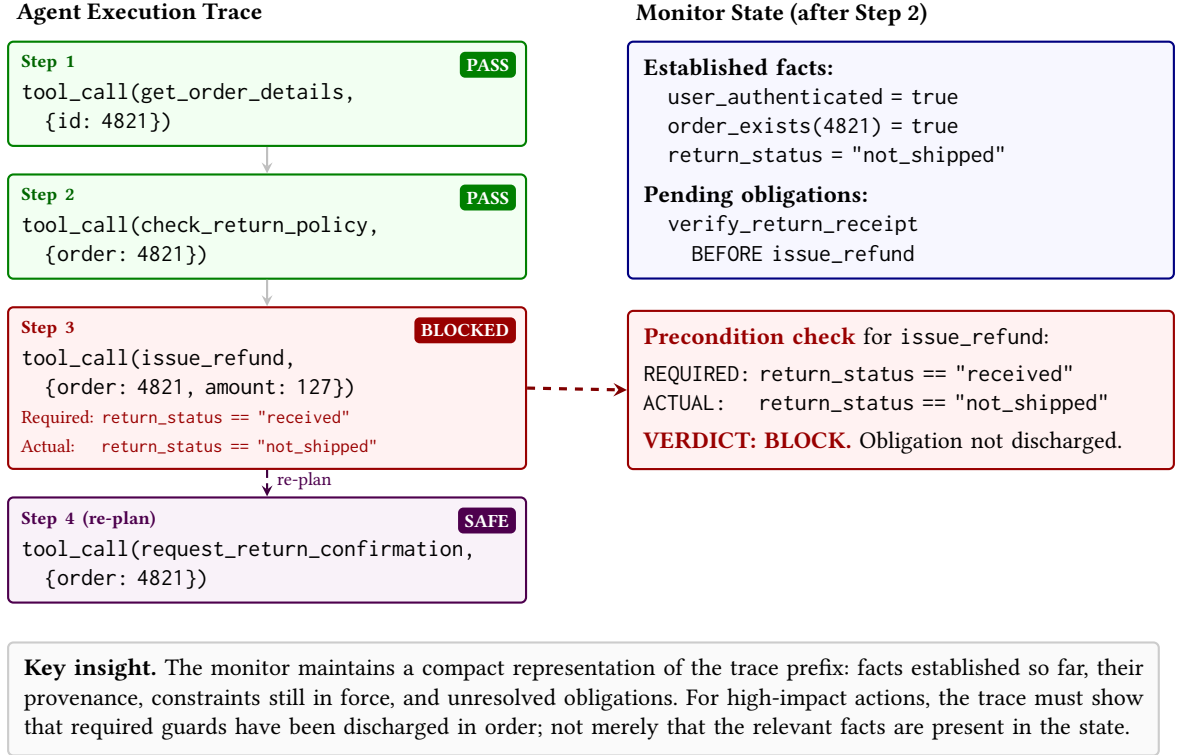
\subsection{Dynamic Analysis}
\label{sec:runtime-analysis}
As AI agents become increasingly complex, weaving user inputs, prompts, tools, models, and memory in nontrivial ways, their reliability hinges on reasoning about behavior during execution.
Static analysis alone is insufficient in these dynamic environments.
Tool outputs are observed incrementally, user constraints may be clarified over multiple turns, memory and retrieved context may conflict with fresh observations, and the external environment may change while the agent is acting.
Dynamic analysis therefore complements static analysis by treating execution as an evolving semantic object, in the spirit of runtime verification, where
properties are checked over observed executions~\citep{leucker2009brief,bauer2011runtime}.
At each step, a runtime monitor reasons over the current trace prefix: the user inputs, model decisions, tool calls, tool responses, confirmations, retrieved evidence, and memory accesses observed so far. 
This prefix induces the monitor's view of the task state and determines which actions are 
justified.
We explain these ideas primarily through tool calls, but the same view extends to retrieval, memory, confirmations, delegation, and environment observations~\citep{yao2022react,schick2023toolformer}.

\begin{table*}[t]
\centering
\caption{Analysis techniques for agent reliability, organized by phase. Static analysis checks agent configurations before execution; dynamic analysis monitors behavior during execution.}
\label{tab:analysis-techniques}
\footnotesize
\begin{tabular}{@{}l p{4.2cm} p{5.5cm} p{3.8cm}@{}}
\toprule
\textbf{Phase} & \textbf{Analysis Technique} & \textbf{What It Checks} & \textbf{Artifact / Object} \\
\midrule
\multirow{5}{*}{\rotatebox[origin=c]{90}{\textbf{Static Analysis}}}
 & Prompt testing and analysis
 & Are task obligations explicit? Does the prompt specify when to call tools, stop, escalate, or ask for confirmation?
 & Prompt text \\

 & Tool contract validation
 & Are tool preconditions, effects, and required evidence explicit? Are high-impact actions guarded?
 & Tool schemas \\

 & Cross-artifact consistency
 & Can individually safe artifacts compose into unsafe behavior? Do the prompt, tools, and policies fit together, or do they create gaps where the agent can take an unsafe action?
 & Prompt $\times$ Tools $\times$ Policy \\

 & Information-flow analysis
 &  Can untrusted information influence sensitive actions without validation?
 & Dataflow across artifacts \\

 & Abstract interpretation
 & Can an unsafe action become reachable before its required evidence, authorization, or confirmation is established?
 & Abstract agent state \\

\midrule
\multirow{5}{*}{\rotatebox[origin=c]{90}{\textbf{Dynamic Analysis}}}
 & Runtime state tracking
 & What facts, permissions, constraints, obligations, and provenance are established by the current execution prefix?
 & Execution prefix \\

 & Evidence-grounded checking
 & Before a tool call, are its preconditions justified by observed evidence in the current trace state?
 & Tool call vs.\ trace evidence \\

 & Temporal guard checking
 & Have required guards been discharged in the right order? For example, authenticate $\to$ validate $\to$ confirm $\to$ commit.
 & Ordering over trace events \\

 & Global invariant checking
 & Do accumulated actions over the trace satisfy global properties such as budgets, resource bounds, authorization, and information-flow restrictions?
 & Full trace \\
\bottomrule
\end{tabular}
\end{table*}
The rest of this section decomposes dynamic analysis into a sequence of
prefix-based validation questions. 
First, what state should the runtime monitor maintain
from the execution prefix? 
Given that state, when is a proposed tool call
justified? 
Which observations still count as valid evidence? 
What obligations
must be discharged before high-impact actions? 
What properties must hold across the accumulated trace?
And when a violation occurs, can the execution still be recovered?
\paragraph{Trace Semantics: What State Does a Runtime Monitor Maintain?}
Dynamic analysis runs while the agent executes a task and interacts with its
environment.
Its central object is not an isolated tool call, but the trace prefix that
precedes it. 
At any point in execution, this prefix contains the user constraints, tool responses, retrieved evidence, commitments, permissions,
consumed resources, and failed actions observed so far. 
Together with the agent's policies and tool contracts, this prefix determines
what the agent is justified in believing and which actions it may safely take next.

The runtime state is a compact representation of this prefix. 
It records the facts established so far, their provenance, the constraints still in force, and
the obligations that remain unresolved.
This state is updated as the agent receives user inputs, observes tool responses, retrieves
evidence, makes commitments, or consumes resources.
A concrete implementation may separate evidence tracking from task
tracking: one component records tool responses, retrieved evidence, memory
entries, and provenance, while another tracks goals, constraints, subgoals, task
progress, and resource use.
These components enable the agent to reason about its current state and make informed decisions based on evolving context. 

This abstraction is necessary whenever correctness depends on runtime
conditions.
For example, in a retail task involving several purchases, the monitor may
need to track the cumulative amount spent across tool invocations to ensure that
the agent remains within the user's budget.
Once the runtime monitor has constructed this state, the next question is whether a
proposed tool call is justified by it~\cite{agentspec,barke2026agentrx}.
\paragraph{Evidence-Grounded Tool Use: When Is a Tool Call Justified?}
Runtime state becomes useful when the agent is about to act.
At that point, the monitor must decide whether the proposed tool call is
justified by the execution prefix.
Correctness is not limited to whether an individual tool call is well-formed. 
A tool invocation may satisfy its schema and still be unwarranted by the trace. 
Dynamic analysis checks whether each action is supported by the facts available at that point in the execution, not merely whether the call conforms to a schema~\citep{agentspec,kamath2025enforcing}.
For example, an \texttt{exchange\_delivered\_order\_items} call may contain all
required arguments but still be invalid if the agent has not established that
the item was delivered, that the replacement satisfies the user's constraints,
and that the user has confirmed the exchange.
These requirements are not merely
syntactic properties of the call. 
They are semantic dependencies on prior events
in the trace. 
Similarly, if a user asks for a product with ``clicky switches,'' the
agent must verify that a prior response
has established \texttt{switch\_type = clicky}. 
The relevant question is not whether
the agent mentioned clicky switches, but whether the trace contains trusted
evidence for that claim.
Dynamic analysis therefore asks whether the preconditions of the proposed action
are derivable from the current runtime state.

\paragraph{Provenance and Freshness: Which Observations Still Count?}

Evidence-grounded tool use assumes that the facts in runtime state are still usable when the agent acts.
This assumption can fail.
%
%
Facts in an agent trace have provenance and lifetime: they come from particular sources, are observed at particular times, and may be weakened or invalidated by
later events.
The monitor must therefore decide whether those facts still count as valid evidence.
An order retrieved early in the session may change before the agent acts on it;
a memory entry may be overridden by a fresh user instruction; a retrieved
document may conflict with a tool response; and a failed API call may invalidate
a candidate plan.
Runtime state must therefore track where a fact came from, when it was observed, and whether later events have weakened or invalidated it.
For example, a calendar agent may check a user's availability early in a
conversation.
If it schedules a meeting much later, the earlier availability result may no
longer be reliable.
The monitor should require a fresh availability check before creating the event.
Thus, dynamic analysis must treat trace observations as time-scoped evidence.

\paragraph{Temporal Guards: What Must Happen Before High-Impact Actions?}

Some action preconditions are temporal: they require particular events to occur before an action is allowed.
For high-impact actions with potential side effects, it is not enough for the relevant facts to be present in the runtime state.
The trace must also show that required guards have been discharged in the right order.
These guards capture familiar ordering requirements: authenticate before account
changes, validate recipients before sending confidential information, check
availability before scheduling meetings, run tests before committing code, and
obtain confirmation before submitting an exchange.
For example, an \texttt{exchange\_delivered\_order\_items} call should not be
allowed merely because the item is eligible and the replacement is valid.
The trace must also contain a prior confirmation event, such as
\texttt{confirm\_exchange}, before the exchange is committed.
The monitor treats confirmation as a pending obligation until a matching
\texttt{confirm\_exchange} event appears in the trace.
Temporal guards capture ordering constraints over individual actions, but some correctness
 conditions are not tied to a single guard.
They must be checked over the accumulated trace as a whole.

\paragraph{Global Trace Properties: What Must Hold Across the Execution?}
Temporal guards capture ordering constraints around individual actions, but some correctness conditions only make sense over the accumulated trace.
A sequence of locally valid actions may still violate a global property.
For example, an agent may make several purchases that individually satisfy the tool schema but collectively exceed the user's budget; schedule meetings that are each valid in isolation but overlap in time; or pass private information through a sequence of steps that eventually exposes it to an unauthorized recipient.
Dynamic analysis therefore maintains invariants over the evolving trace state, such as budget limits, resource bounds, compatibility constraints, authorization requirements, and information-flow restrictions.
When these properties can be encoded symbolically, SMT-style checks can be used to test whether the current state remains consistent with the accumulated constraints~\citep{de2008z3}.
For instance, a shopping agent may add items one at a time, each satisfying the user's local preferences, while the running total eventually exceeds the user's budget.
The runtime state must therefore maintain the accumulated cost and reject any next purchase that violates the budget constraint.

\paragraph{Recoverability and Feedback: What Happens After a Violation?}
A runtime violation need not mean that the execution has failed outright.
Some violations are wrong turns: the agent relied on stale data, skipped a confirmation, chose an incompatible item, or followed a plan invalidated by a failed tool call.
In these cases, the monitor can still steer the execution back onto a valid path by blocking the next action, asking for missing evidence, refreshing state, or triggering replanning.
Other violations cross a point of no return.
Once an email has been sent, an order submitted, private information disclosed, or a destructive operation committed, later steps cannot simply repair the trace.
This distinction matters because recoverable violations call for
repair or replanning, while unrecoverable violations call for aborting, escalating, or restarting.
%
Violations are also useful beyond the current execution.
Repeated failures show where the agent's specification is incomplete. A missing confirmation check, a stale memory lookup, or an unsafe path to a high-impact tool is not just an execution error; it is evidence that the prompt, tool interface or memory policy failed to encode a needed constraint.
In this way, runtime feedback can support specification mining: successful executions, failed traces, documentation, and human feedback can be used to propose candidate preconditions, temporal guards, data dependencies, and invariants for a domain.
These mined specifications are not ground truth, but they provide candidates that can be reviewed and promoted into prompts, tool contracts, runtime monitors, and regression tests.
Runtime monitoring, recovery, and specification revision together form a feedback loop related to the Monitor--Analyze--Plan--Execute over Knowledge (MAPE-K) architecture from autonomic computing~\cite{kephart2003autonomic} and its broader development in self-adaptive systems~\cite{brun2009engineering,delemos2013roadmap}.
In our setting, the knowledge component includes an incomplete
specification that may itself be revised in response to observed
failures.

\section{Conclusion}
\label{sec:conclusion}

In this essay, we argued for the conception of AI agents as software systems from a programming languages perspective: structured, composable units with clear functional roles that can be analyzed, tested, and verified. We then described an agenda involving the expression of clear specifications written over well-defined formal models of agentic executions and involving new and complex specification logics. We further described a variety of static and dynamic enforcement methodologies for various classes of specifications, and discussed the nuances involved in designing such algorithms. Our hope is that this essay acts as a sound starting point for the community to develop a rich agenda on reliability for AI agents.
\bibliographystyle{ACM-Reference-Format}
\bibliography{refs}
\end{document}